# SiN$_x$ RRAMs performance with different stoichiometries

A.E. Mavropoulos [a,b,*], G. Pissanos [a], N. Vasileiadis [a,b], P. Normand [a], G. Ch. Sirakoulis [b], P. Dimitrakis [a,*]

[a] *Institute of Nanoscience and Nanotechnology, NCSR "Demokritos", Ag. Paraskevi 15341, Greece*
[b] *Department of Electrical and Computer Engineering, Democritus University of Thrace, Xanthi 67100, Greece*

**Abstract**

The microstructure of SiN$_x$ is strongly affected by its stoichiometry, x. The stoichiometry of SiN$_x$ thin films can be modified by adjusting the gas flow rates during LPCVD deposition. The deficiency or excess of Si atoms enhance the formation of defects such as nitrogen vacancies, silicon dangling bonds etc., and thus can enable performance tuning of the resulting MIS RRAM devices. DC electrical characterization, impedance spectroscopy and constant voltage stress measurements were carried out to investigate the properties of non-stoichiometric silicon nitride films as resistive switching material. The average SET time for each device was measured by applying voltage ramps. Improvement in the SET/RESET voltages and SET time is observed. Finally, the stoichiometric film exhibits the lowest breakdown acceleration factor, while the Si-rich film the highest.

## 1. Introduction

Silicon nitride (SiN$_x$) has been widely used in charge-trapping (CT) nonvolatile memories for many years [1, 2]. Intrinsic defects with different configurations (i.e. nitrogen vacancies and Si dangling bonds) are formed in the material due to the nitrogen deficiency that is caused by the thermodynamics of the deposition method [3]. Functional Metal-Insulator-Metal (MIM) and Metal-Insulator-Semiconductor (MIS) resistive memory cells with LPCVD Silicon Nitride as active material, which was found to exhibit competitive resistance switching properties, have been demonstrated [4], [5], [6]. SiN$_x$ RRAMs show great potential for memory cell scaling [7], neuromorphic, in-memory computing [4], [8], [9] and security applications by creating true random number generators [10], [11]. In most publications SiN$_x$ is stoichiometric (x = 1.33). In this work different [N]/[Si] ratios are explored, with the goal to improve the resistance switching characteristics. In our previous work we have discussed that SiN$_x$ RRAMs are Valance-Charge (VCM) and the conductive filaments are formed from nitrogen vacancies [12]. Changing the stoichiometry results in different defect concentrations in the material, thus affecting parameters such as the SET/RESET voltage, the resistance switching speed and the dielectric strength.

## 2. Experimental

Three heavily Phosphorus-doped SOI (100nm Si and 200nm BOX) 2cm × 2cm pieces were used as substrates for the RRAMs. The doping was achieved by ion implantation, with a dose of $1\times10^{15}$cm$^{-2}$ and 40keV, through 20nm of SiO$_2$ (hard mask) and post-implantation annealing at 1050°C for 20s. SiN$_x$ layers were deposited using LPCVD and the flow rates of dichlorosilane (DCS) and ammonia were set according to Table 1, while the deposition time and temperature, 100s and 800°C respectively, were the same for all depositions. This lead to the creation of a stoichiometric sample (S1) and two non-stoichiometric (S11, S12). The top electrode (TE) consists of 30nm Cu and 30nm Pt capping layer to prevent the oxidation of copper, and the bottom electrode (BE) of 100nm Al. Three more samples are also fabricated on undoped SOI pieces, for capacitance and ellipsometry measurements. The schematic of the fabricated devices is presented in Figure 1.

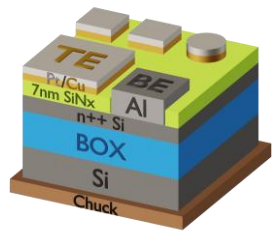

**Figure 1** Device schematic

## 3. Results and discussion

Initially, the layer thickness of the undoped samples is measured using ellipsometry and the results are shown in Table 1. Higher DCS flow rates lead to thicker layers, because the number of Si atoms provided for the chemical reaction is increased. The deposition time needs to be adjusted to achieve the same thickness on all samples. The index of refraction for S11 is lower than S1 and for S12 higher. According to [13] higher concentrations of Si lead to a higher index of refraction, so we can assume that S12 is silicon rich and S11 nitrogen rich, compared to S1.

**Table 1** *Fabricated samples characteristics*

| Sample | DCS (sccm) | NH$_3$ (sccm) | Thickness (nm) | SET V/E (V/MV×cm$^{-1}$) | RESET V/E (V/ MV×cm$^{-1}$) |
|---|---|---|---|---|---|
| S1 (Stoich.) | 20 | 60 | 8.8 | 3.4/3.9 | -2.3/-2.6 |
| S11 (N-rich) | 20 | 20 | 7.2 | 3.6/5.0 | -1.9/-2.6 |
| S12 (Si-rich) | 60 | 60 | 11.5 | 5.9/5.1 | -1.9/1.7 |

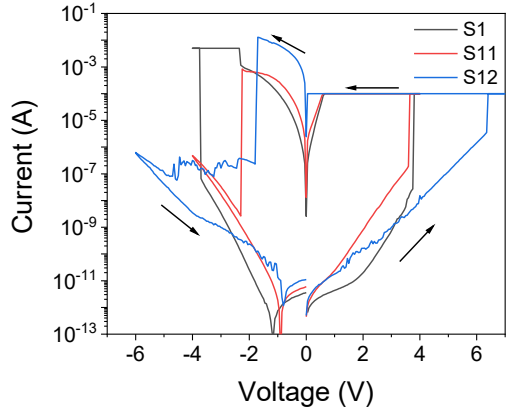

**Figure 2** I-V sweeps with $I_{CC}$ = 100 μA

HP4155 and Tektronix 4200A were used for DC electrical characterization of the RRAM cells and impedance measurements were carried out using HP4284 and Zurich Instruments MFIA. For all the measurements the bottom electrode was grounded, and the voltage was applied on the top electrode. A wafer prober was used to make connections to the samples, which were measured at room temperature.

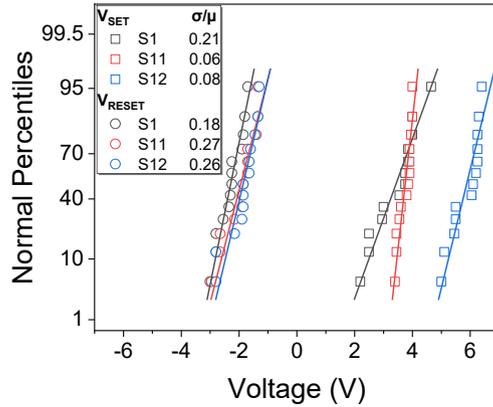

**Figure 3** Statistical analysis of SET and RESET voltages

I-V sweeps were performed on all samples with the implanted substrate using a 100μA compliance current ($I_{CC}$) and the results are shown in Figure 2. It becomes evident that for stoichiometric and rich in N $SiN_x$ films the peak current during the RESET is lower, indicating self-compliance characteristics. These two stoichiometries result in a low RC time constant, limiting the current overshoot [14].

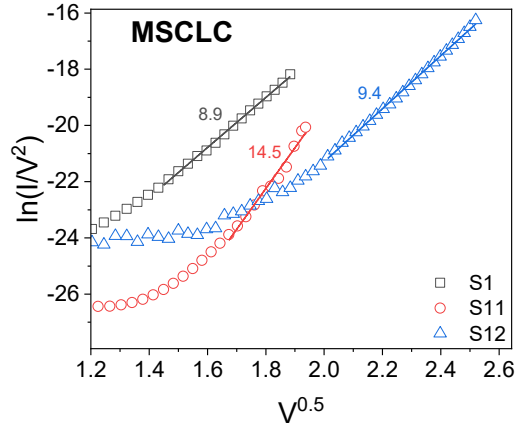

**Figure 4** Analysis of the I-V plots during SET according to the MSCLC conduction mechanism

Moreover, the SET voltage is higher for S12. This is confirmed by statistically analyzing the SET and RESET voltages, as seen in Figure 3 and Table 1. This is because of the higher thickness of the deposited layer on S12. The SET voltages in general seem to be more dependent on the thickness and not the composition of $SiN_x$. If the electric field is calculated, it becomes evident that moving away from the stoichiometric $SiN_x$ leads to an increased SET electric field. On the other hand, the RESET voltages are very similar for all samples, however the electric field is lower for the Si-rich. In addition, the coefficient of variation σ/μ, which is defined as the ratio of the standard deviation σ over the mean value μ is calculated. For the SET it decreases when we move away from the stoichiometric $SiN_x$, meaning that there is less variation in the SET voltages.

An analysis of I-V curves revealed that the modified space charge limited conduction mechanism (MSCLC) [15] fits the best during SET, as shown in Figure 4. This mechanism fits very well the I-V curves of Si-doped $SiN_x$ RRAMs in our previous work [12]. This conduction mechanism is a combination of the Poole-Frenkel and the Space Charge Limited (SCLC) mechanisms. This means the conduction is both trap limited and bulk limited. The following equation is used to describe the previously mentioned conduction mechanism:

$$I = \alpha V^2 e^{\beta\sqrt{V}} \qquad (1)$$

$$\alpha = \frac{9\varepsilon\varepsilon_0 \mu N_C A}{8 d^3 N_T} e^{-q\varphi_T/k_B T} \qquad (2)$$

$$\beta = \frac{q}{k_B T}\sqrt{\frac{q}{\pi\varepsilon\varepsilon_0 d}} \qquad (3)$$

where A, d, ε, $N_C$, μ are the device area, film thickness, dielectric constant, density of states in the conduction band of the dielectric and the electronic drift mobility respectively and all the other symbols have their usual meaning. Also, $\varphi_T$ and $N_T$ are the trap energy and concentration respectively.

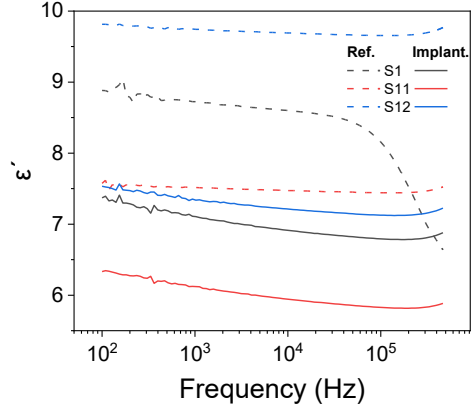

**Figure 5** Dielectric constant for all $SiN_x$ on both implanted and non-implanted substrates

Impedance spectroscopy measurements were applied on all devices on both the HRS and LRS, as well as pristine samples without any prior measurements. An AC small signal of 25 mV with a +0.1 V DC bias was used. From the pristine devices it is possible to calculate the dielectric constant ε', which is portrayed in Figure 5. ε' increases as the $SiN_x$ gets richer in Si (it gets closer to that of pure Si: 11.7) and decreases when it gets richer in N. The equations

$$\varepsilon(\omega) = \frac{1}{i\omega C_0 Z(\omega)} \qquad (4)$$

and

$$\sigma(\omega) = i\omega\varepsilon_0\varepsilon(\omega) \qquad (5)$$

were used to calculate the dielectric constant and the conductance from the impedance measurements, where ω the angular frequency, $\varepsilon_0$ the permittivity of free space and $C_0$ the geometrical capacitance ($C_0 = A\varepsilon_0/d$, d: thickness of the dielectric, A: area of the dielectric) [16].

**Table 2** *Modelling parameters for Nyquist plots at LRS.*

| Sample | $C_P$ (pF) | $R_P$ (kΩ) |
|---|---|---|
| S1 | 274.2 | 42.2 |
| S11 | 294.1 | 17.7 |
| S12 | 214.2 | 17.0 |

The Nyquist plots for the Low Resistance State (LRS) form a semicircle (Figure 6(a)), indicative of the response of an $R_s$-($R_p$||$C_p$) circuit where $R_p$ is the resistance of the conductive paths in $SiN_x$, $C_p$ is the capacitive response of the remaining (unswitched) insulating volume in the examined MIS capacitor and $R_s$ the series resistance [14], [17]. The calculated modelling parameters are shown in Table 2.

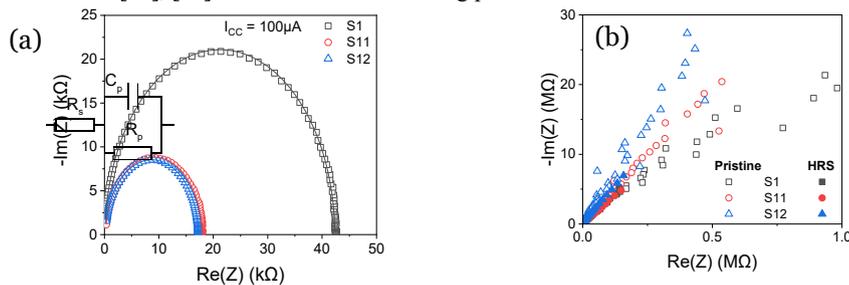

**Figure 6** (a) Nyquist plots for the LRS. Continious lines indicate the $R_s$-($R_p$||$C_p$) model and (b) Nyquist plots for pristine samples and at a HRS

The Nyquist plots for the HRS form a line, which means a Warburg impedance is dominant. They are also similar to the ones for pristine samples as seen in Figure 6(b), which means the devices return close to their original state during the RESET, but with a slightly lower resistance, which could be due to the conductive filament not fully dissolving.

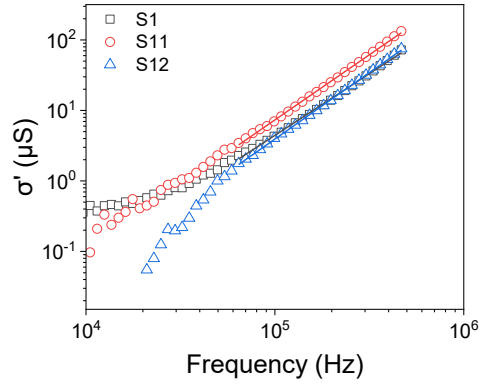

**Figure 7** AC conductivity measurements

The conductivity was also measured at LRS and by removing the DC part, the AC conductivity is plotted in Figure 7 and the results suggest that σ' varies as $\sim f^s$. By performing a linear fit on these plots, a slope of 1.87 was calculated for S1, 1.91 for S11 and 1.94 for S12. These values of s are close to 2 and suggest, according to [18], that the conduction is mainly due to trap-to-trap tunneling mechanisms.

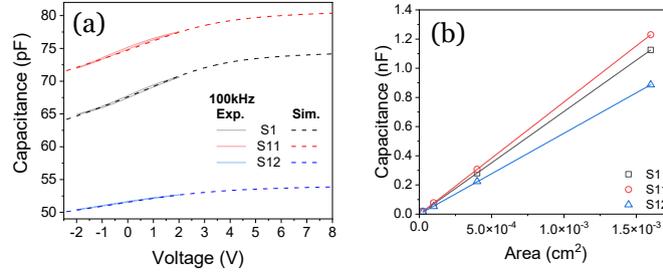

**Figure 8** (a) Experimental capacitance measured at 100kHz of a 100μm × 100μm device for each sample compared to simulated capacitance and (b) capacitance vs RRAM area

The capacitance of all samples for MIS capacitors with different areas was measured using a 25 mV AC signal, while sweeping the DC bias from -3V to +3V (Figure 8(a)). In order to better calculate the accumulation capacitance, the C-V curves were simulated using the Boise State University Band Diagram Program [19]. The capacitance during the accumulation was plotted versus the area in Figure 8(b) and a linear correlation is observed. This means that the thickness of each $SiN_x$ can be calculated. For S1 it is found to be 9.0 nm, for S11 7.2 nm and for S12 12.6 nm, quite similar to the results of the ellipsometer.

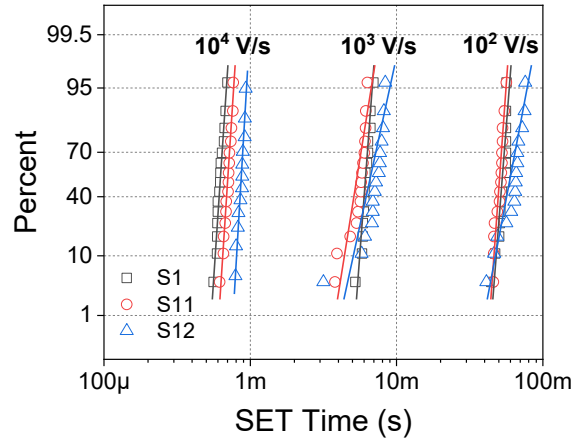

**Figure 9** Statistical analysis of the average time for SET with different ramp rates

Voltage ramps with three different ramp rates, $10^2$ V/s, $10^3$ V/s and $10^4$ V/s, where applied and the time to switch to the LRS was investigated. A clear trend appears after statistically analyzing the SET times (Figure 9), with N-rich samples switching quicker and Si-rich slower than the stoichiometric. The mean SET times are presented in Table 3.

**Table 3** Mean SET time for all samples at different ramp rates

| Ramp (V/s) | S1 (ms) | S11 (ms) | S12 (ms) |
|---|---|---|---|
| $10^2$ | 53.0 | 50.8 | 62.6 |
| $10^3$ | 6.16 | 5.53 | 7.03 |
| $10^4$ | 0.66 | 0.63 | 0.82 |

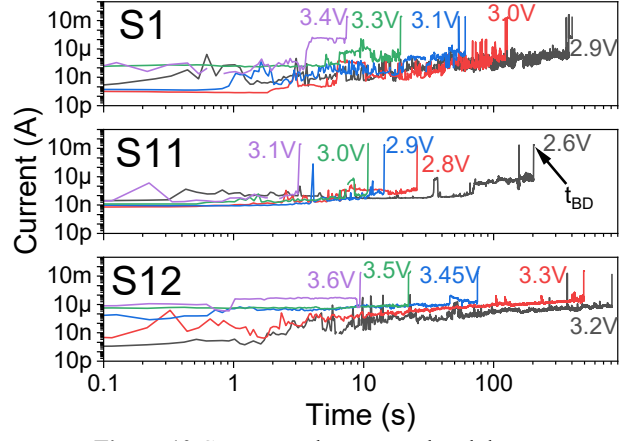

**Figure 10** Constant voltage stress breakdown

Finally, time dependent dielectric breakdown tests on pristine devices were performed. In these experiments, a constant voltage was applied on a device, which is in the HRS, and the current was measured every 100ms. The current evolution through time at various constant voltages is shown in Figure 10. In Figure 11, the $t_{BD}$ dependence on the $1/E_{Stress}$ is demonstrated. In case of defect related breakdown, according to [20]

$$t_{BD} = Ae^{B(d-\Delta d)/V_{Stress}} = Ae^{B(1-\frac{\Delta d}{d})/E_{Stress}} \qquad (6)$$

where $V_{Stress}$ ($E_{stress}=V_{stress}/d$) is the stress voltage (field), d is the thickness of the dielectric, $\Delta d$ is the effective dielectric thinning due to the defects formed inside the $SiN_x$ and B is a fitting parameter. The slope of linear least-square fitting in Figure 11 is called the electrical field acceleration factor and is equal to

$$\frac{dln(t_{BD})}{d(1/E_{stress})} = B(1-\frac{\Delta d}{d}) \qquad (7)$$

This slope increases as the $SiN_x$ gets richer in Si, meaning smaller changes in the electric field are required for the breakdown time to decrease. The stoichiometric $SiN_x$ requires about 3.5MV/cm to break down and has the smallest electric field acceleration factor, while the N-rich 4.3MV/cm and the Si-rich 2.7MV/cm. The existence of more traps in the Si-rich sample can explain the lower fields required to break down. However, the $V_{Stress}$ is higher for the Si-rich as seen in Figure 10. This can be the result of the thicker films of the Si-rich samples. The Si-rich samples are thicker because the Si atoms have larger volume compared to N atoms, and thus the effective electric field is lower, resulting in longer breakdown times.

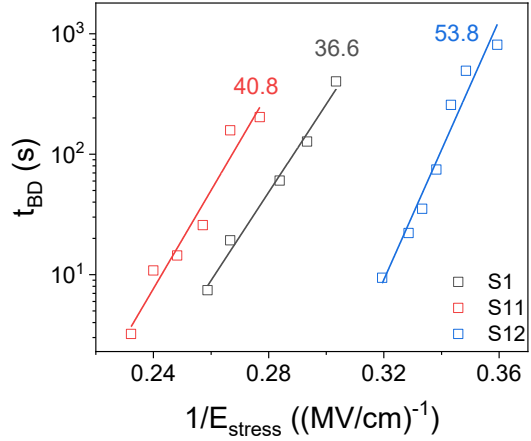

**Figure 11** Time of breakdown vs $1/E_{Stress}$

## 4. Conclusions

In conclusion, it is possible to change the ratio of N and Si of a $SiN_x$ layer deposited by LPCVD by modifying the flow rates of the DCS and $NH_3$ gases. Stoichiometric, N-rich and Si-rich $SiN_x$ layers were deposited, with ellipsometry and impedance spectroscopy measurements revealing that the microstructure plays an important role in the characteristics of the films, affecting the thickness and the dielectric constant. The dielectric constant increases as $SiN_x$ gets more Si-rich. Stoichiometric and N-rich $SiN_x$ RRAMs demonstrate self-compliance characteristics during the RESET. Moreover, $V_{SET}/V_{RESET}$ mainly depend on the thickness and not on x and the MSCLC conduction mechanism fits the best during the SET. In addition, the SET time decreases as the $SiN_x$ gets more N-rich. Finally, $log(t_{BD}) - 1/E_{stress}$ have linear dependance, with the stoichiometric samples having the smallest electric field acceleration factor. The Si-rich RRAMs require lower electric fields to break down, due to the existence of more defects. For future work, the deposition time should be adjusted to obtain films of the same thickness for a better comparison of the materials themselves. In addition, more gas ratios will be tested to better fine tune the conduction properties of the $SiN_x$.

**Declaration of Competing Interest**

The authors declare the following financial interests/personal relationships which may be considered as potential competing interests: This work was supported in part by the research project "LIMA-chip" (Proj.No. 2748) which is funded by the Hellenic Foundation of Research and Innovation (HFRI).

**Acknowledgements**

This work was financially supported by the research project "LIMA-chip" (Proj.No. 2748) of the Hellenic Foundation of Research and Innovation (HFRI).